\documentstyle[12pt]{article}
\begin{document}
\vskip 4 cm
\begin{center}
{\Large \bf
ON THE EQUIVALENCE THEOREM IN THE $\chi PT$ DESCRIPTION OF THE SYMMETRY
BREAKING SECTOR OF THE STANDARD MODEL}
\vskip 1cm
{\large \bf A. Dobado} \footnote{On leave of absence from  Departamento
de F\'{\i}sica Te\'orica, Universidad Complutense de Madrid, 28040 Madrid,
Spain } \footnote{E-mail: dobado@cernvm.cern.ch}    \\
{\large Department of Physics, Stanford University\\
Stanford, CA 94305-4060, USA\\
 and  \\}
{\large \bf J. R. Pel\'aez} \footnote {E-mail: pelaez@fis.ucm.es }\\
{\large Departamento de F\'{\i}sica Te\'orica II \\
 Universidad Complutense de Madrid\\
 28040 Madrid, Spain \\}
\vskip 1cm
 SU-ITP-93-33, hep-ph/9401202 \\
November 1993 \\
Extended version \\
April 1994\\
\vskip 1cm
\begin{abstract}
We develop an alternative formulation of the symmetry breaking sector of the
 Standard Model  as a gauged non-linear sigma model (NLSM) following the
philosophy of the
 Chiral lagrangian approach, which is the only compatible with all the
experimental and theoretical  constraints. We derive the BRS symmetry of the
model and the corresponding quantum lagrangian, which
 is a generalization of the standard  Faddeev-Popov method, in a way which is
covariant with respect to
 the reparametrizations of the coset space of the NLSM. Then we use the BRS
invariance of the
 quantum lagrangian to state the Equivalence Theorem for the renormalized
$S$-matrix elements calculated as a chiral expansion.
\end{abstract}
\end{center}
\newpage


\section{Introduction}

In the last years, a great deal of work has been devoted to the so called
Chiral Perturbation Theory ($\chi PT$) description of the symmetry breaking
sector of the Standard Model (SM). Originally, $\chi PT$ [1] was developed
for the description of hadronic physics at low energies where perturbative
Quantum Cromodynamics (QCD) cannot be directly applied. $\chi PT$ is based in
the approximate $SU(2)_L\times SU(2)_R$ symmetry of the hadronic
 interactions which is supposed to be spontaneously broken to the $SU(2)_{L+R}$
(isospin) subgroup. The corresponding Goldstone bosons (GB) are identified with
the pions and their dynamics is obtained as a derivative expansion with
couplings that have to be fitted from
 the experiments but that one could, at least in principle, compute from the
underlying theory, i.e. QCD.

In the case of the symmetry breaking sector of the SM things are quite
different but still we have some analogies with the low-energy hadron dynamics.
Despite the very accurate
 precision measurements of the electroweak parameters recently performed at the
Large
Electron-Positron Collider (LEP), it is very few what is really known about
the physical mechanism responsible for the breaking of the $SU(2)_L \times
U(1)_{Y}$ gauge symmetry to the electromagnetic group $U(1)_{em}$ (see [2] for
pedagogical review). Basically,   what we know is the following: There
 must be a system coupled to the SM with some global symmetry $G$ which is
spontaneously
 broken through some undetermined mechanism to some subgroup $H$. This global
symmetry breaking drives the spontaneous breaking of the gauge symmetry
$SU(2)_L \times U(1)_{Y}$ to $U(1)_{em}$ of the SM responsible for the
$W^{\pm}$ and $Z$ masses. Therefore, the only established modes of that system
are the three  GB corresponding to the global symmetry breaking $G$ to $H$.
Finally,
 these degrees of freedom will appear related through the Higgs mechanism to
the longitudinal components of the massive vector bosons.

Apart from these very
general considerations we also have some quantitative information about
the symmetry breaking sector of the SM like the value of the dimensional
constant $v \simeq 250 GeV$ appearing in the low-energy matrix element
between the GB and the gauge currents which can be obtained (using the gauge
symmetry of the SM)  from the muon lifetime. In addition we have the
values of the gauge boson masses and the $\rho$ parameter which is
known to be very close to one. It has
been argued that the presence in the symmetry breaking sector of the SM of the
 so called custodial symmetry $SU(2)_{L+R}$
[3] as a subgroup of $H$ yields
$\rho=1$ when considering only the contribution from this sector. The
corrections coming from the coupling with the gauge bosons
 are in general small and compatible with the present experimental value of
$\rho$.

Basically, this is all we really know about the symmetry breaking sector of the
SM. Of course, in the literature one can find a huge number of proposals for
the
description of this sector, being the most well known the minimal SM (MSM) with
a Higgs doublet and a potential defined ad hoc to produce the symmetry
breaking, Technicolor
 (TC) [4] where one uses a similar mechanism to that producing the rupture
of the chiral symmetry in QCD,
 and Supersymmetry (SUSY) (see [5] for a review) where, apart from the MSM
fields, we have at least an extra
 doublet of Higgs fields and all their corresponding SUSY partners.

Having no idea of the method used by nature to give masses to the electroweak
bosons, it seems to be quite reasonable to look for a
model independent description of the GB low energy dynamics using nothing more
than the well known facts discussed above. It is here then where the use of the
$\chi PT$ methods become extremely useful. The first application of $\chi
PT$ including the effect of GB loops to this context [6] dealt with the
description of the elastic scattering of the gauge
bosons longitudinal components, but more recently it has also been used to
parametrize the precision tests of the SM coming from LEP [7].

Concerning the first kind of application, the so called Equivalence
Theorem (ET) plays an essential role. This theorem states that, for
renormalizable or $R_{\xi}$ gauges and at high energies compared with
 the gauge boson mass, the $S$-matrix elements for the GB are given by the
corresponding $S$-matrix elements for the longitudinal components of the gauge
boson (in the following we will call them generically $W_L$).  The ET is
usually assumed in many phenomenological descriptions of the elastic
$W_LW_L$ scattering independently of the nature of the symmetry breaking sector
of the SM. However, despite there are many heuristic
arguments suggesting that some version of this
 theorem should apply in the general case, not only there is no proof of this
fact but even there are some indications pointing that the concrete form of the
theorem depends on the way the computations of the $S$-matrix are done. The
first version of the theorem was given by Cornwall,  Levin and
 Tiktopoulos and also by Vayonakis [8] at the tree level and it was extended at
any order in the Feynman 't Hooft gauge by Lee, Quigg and Thacker for the
 case of one leg [9]. Later on, Chanowitz and Gaillard gave a complete and
very systematic proof of the theorem valid at any order and for any number of
legs [10] using the Ward-Slavnov-Taylor identities derived from the
Becchi-Rouet-Stora (BRS) symmetry [11] of the SM. Further, this demonstration
 was slightly simplified by Gounaris, Kogerler and Neufeld [12]. Later, it was
realized by Yao and Yuang [13] that renormalization affects differently the GB
and the gauge bosons, and so produces corrections to the ET which in principle
depend on the
 renormalization scheme and the gauge choice. Recently some articles have
appeared trying to clarify this and other issues related with the ET [14].

In any case, it must be stressed that most of the work done on the ET refers
exclusively to the MSM, and therefore, there are no solid reasons to apply this
theorem when other symmetry breaking sectors or mechanisms are considered. In
this paper we  will try to fill this gap by obtaining the version of the
ET that applies when the symmetry breaking sector of the
SM is described by a general chiral lagrangian. The path
we will pursuit to that end is
roughly the following: First we will develop in detail the most general
description of the GB dynamics compatible with the real information we have on
the symmetry breaking sector of the SM. This will be done as in $\chi PT$ i.e.
using a $SU(2)_L \times U(1)_Y$ gauged non-linear sigma model (NLSM)
derivative expansion
 based on the coset $G/H$ so that all the information
 (or better the lack of information) about the GB dynamics will be parametrized
by the couplings of the higher derivative terms. We will show how it is
possible
to do that independently of the coordinates used to parametrize the coset
space $G/H$ so that the model
  only depends on $G$ and $H$. Then we will find the BRS transformations
corresponding to the $SU(2)_L \times U(1)_Y$ symmetry of the gauged NLSM and we
will obtain a BRS invariant quantized lagrangian using a generalization of the
standard Faddeev-Popov method which gives rise to a well defined propagator
 for the gauge bosons, and therefore is appropriate for doing perturbative
computations. The next step is the introduction
of the renormalized lagrangian which is also BRS invariant due to the
absence of
anomalies in the SM. Then, as BRS invariance is a key ingredient of the formal
Chanowitz-Gaillard proof of the ET we will be able to follow many of their
steps
also in our case. However, as we will see, some important differences arise due
to the renormalization factors and the peculiarities of the $\chi PT$
description of the SM symmetry breaking sector.

Thus, the plan of the paper goes as follows. In sec.2 we define, according to
the
present experimental information, the NLSM based on the coset $G/H$ describing
the
dynamics of the GB that appear in the SM symmetry breaking sector,
discussing its main properties both at the classical and at the quantum level.
In
sec.3 we introduce the $SU(2)_L \times U(1)_Y$
 gauge degrees of freedom by gauging in the directions of the appropriate $G$
killing vectors to find the gauged NLSM which describes the interactions of the
GB
and the gauge bosons. Starting from the gauge transformations of the gauged
NLSM we introduce in sec.4 the corresponding BRS and anti-BRS  transformations
building up the (anti-) BRS invariant
 quantum lagrangian appropriate for perturbative computations in $R_{\xi}$
gauges. In sec.5 we introduce the renormalized lagrangian and the  renormalized
BRS and anti-BRS  transformations which leave it invariant when it is written
in
terms of renormalized fields. Then, in sec.6, and following  the steps of the
Chanowitz-Gaillard proof, we make use of the BRS invariance to obtain
Ward-Slavnov-Taylor Identities between renormalized Green functions containing
gauge bosons and GB. In sec.7 we translate these relations to the $S$-matrix
elements obtained through the chiral expansion. Section 8 is devoted to discuss
the real meaning of our result and the existence of an energy region where
$\chi PT$ and the ET can be safely used simultaneously. Finally, in sec.9 we
briefly review the main results of this paper.

\section{The dynamics of the GB and its geometrical interpretation}

The most general description of the GB dynamics compatible with the information
we have on this sector of the SM can be obtained as follows. First we turn
off the gauge fields ($g=g'=0$). The rest of the system must have some global
symmetry $G$ which is spontaneously broken to some subgroup $H$ through some
unknown mechanism. We will represent by $T^a$ the generic hermitian $G$
generators with $a=1,2,...,g=dim G$. They can also be written as $T^a=H^a$ for
$a=1,2...,h=dim H$ and $T^a=X^a$ for $a=h+1,h+2,..., g$ where the $H^a$ are the
generators of $H$. We will assume $G$ to be compact and the coset space $K=G/H$
to be symmetric. This means that the commutator of two $X$-like generators
will be a linear combination of $H$-like generators only.

 It is possible then to
define an involutive automorfism into $G$ as follows. Let $u \in G$ so that we
can write $u=\exp i(\alpha_aH_a+\beta_aX_a)$, and the result of the automorfism
acting on $u$, let us call it $u'$, is obtained from $u$ by changing the sign
of the parameters corresponding to the $X$-like generators, i e. $u'=\exp
i(\alpha_aH_a-\beta_aX_a)$. So defined, the automorfism is obviously
involutive and it has $H$ as the maximal invariant subgroup. From the physical
point of view, the meaning of this automorfism is the parity transformation
which seems to play a central role in the electroweak interactions.

According to the Nambu-Goldstone theorem, the GB fields take values on the
coset
space $K=G/H$, so that there are just $k=dim K=dim G-dim H=g-h$ of them. In
order to parametrize the GB fields we need some coordinates
$\omega^\alpha$ with
 $\alpha=1,2,...,k$ on the coset $K$. Then the group $G$ can be understood as
the isometry group of the coset $K$ when it is equipped with metrics built up
as follows: Let $p$ be some point of $K$ with coordinates $\omega^\alpha$ and
$u=1+iT^a\epsilon^a$ some small $G$ transformation. Then $u$ maps the point $p$
in another point $p'$ of coordinates
$\omega'^\alpha=\omega^{\alpha}+\delta \omega^{\alpha}$ so that we can write:
\begin{equation} \delta \omega^{\alpha}=\xi^{\alpha}_{\;a}(\omega)
\epsilon^a \end{equation} This relation precisely defines how the $G$
symmetry is realized non-linearly by the GB fields $\omega^\alpha(x)$. From the
geometrical point of view, $\xi^a =\xi^{\alpha}_{\;a}\partial/\partial
\omega^\alpha$ can be interpreted as the killing vectors corresponding to the
$G$ transformations acting on $K$. Now it is possible to define the $K$ metrics
through the vielbein $e_a=e^{\alpha}_{\;a} \partial/\partial \omega^{\alpha}$
with $e^{\alpha}_{\;a}=\xi^{\alpha}_{\;a+h}$ for $a=1,2,...,k$ i.e. the
vielbein is just the set of killing vectors corresponding to the $k$ broken
generators. The $K$ metrics $g_{\alpha\beta}$ is defined as the inverse of
$g^{\alpha\beta}$ where:
\begin{equation}
g^{\alpha\beta}=e^{\alpha}_{\;a}e^{\beta a}
\end{equation}
 with $a=1,2,...,k$.
In the following, $\alpha,\beta,...$ will refer to curved coordinates on $K$
and $a,b,...$ to flat or tangent space indices. These indices
 are therefore raised and lowered with the tensors $g_{\alpha\beta}$
and $\delta_{ab}$ respectively. Now it is not difficult to show that the $G$
transformations are isometries of this metric, i.e.
\begin{equation}
g'_{\alpha\beta}(\omega)=g_{\alpha\beta}(\omega)
\end{equation}
 so that, the metric functional dependence of the coordinates on $K$ does not
 change after the $G$ transformation.

With the above defined metric on $K$ it is possible to introduce the GB
dynamics
 intrinsically, or in other words, independently of the choice of the
coordinates on the coset space $K$, and therefore, only dependent on $G$ and
$H$.
 The most general $G$ invariant lagrangian describing the low-energy dynamics
of the GB can be organized by taking into account the number of derivatives of
the different terms which must be covariant not only in the space-time sense
but also in the $K$ sense. For example, the first term has two derivatives and
can be written as:
\begin{equation}
 {\cal L}_0=\frac{1}{2}g_{\alpha\beta}(\omega)\partial_{\mu}\omega^{\alpha}
\partial^{\mu}\omega^{\beta}
\end{equation}
 The form of this lagrangian does not
depend on the coordinates chosen on $K$ since it is $K$ covariant, and it is
$G$ invariant too as can be easily checked using eq.3. Note also that the
lagrangian not only contains the kinetic term but also all the infinite
interaction terms that can be
obtained by expanding the $K$ metrics $g_{\alpha\beta}$ in powers of
$\omega^{\alpha}$ with an arbitrary (even) high number of GB fields .

 Any other term that could be added to ${\cal L}_0$ to write the most general
$K$ covariant and $G$ invariant lagrangian will have more derivatives so that
we just write:
\begin{eqnarray}
 {\cal L}&=&{\cal L}_0+
c_1(g_{\alpha\beta}(\omega)\partial_{\mu}\omega^{\alpha}
\partial^{\mu}\omega^{\beta})^2+
c_2(g_{\alpha\beta}(\omega)\partial_{\mu}\omega^{\alpha}
\partial^{\nu}\omega^{\beta})^2\\ \nonumber &+&higher \; derivative \; terms
\end{eqnarray}
 and so on. The constants appearing in these terms make it possible
to parametrize systematically the GB dynamics of any theory with
spontaneous symmetry breaking pattern from $G$ to $H$.

Once the classical lagrangian has been defined we can go to the quantum theory
by using standard path integral methods. The only subtlety that we have to take
 into account in this case is the following: Any quantum theory is not
uniquely defined by the classical lagrangian but one also has to define a
measure in the corresponding functional field space. Therefore, in order to
have a $G$ invariant and $K$ covariant quantum theory not only do we need a
lagrangian  but also a measure with this properties. Then, as it has
 been discussed in the literature (see for instance [15]), the proper
definition of the generating functional is:
 \begin{equation}
e^{iW[J]}=\int[d\omega\sqrt{g}]e^{i\int d^nx({\cal L}(\omega,\partial^2 \omega,
\partial^4 \omega,...)+J_a\Gamma^{a})}
\end{equation}
 where $g$ is the
determinant of the $K$ metrics, $n$ is the space-time dimension and
$\Gamma^{a}$ is defined as follows: Given the point $p$ of $K$ with coordinates
$\omega^ \alpha$ and some other point $0$ selected as the origin (the
classical vacuum) with coordinates $\omega^ \alpha=0$ we consider the geodesic
curve joining $0$ and $p$ (we assume both points are close enough for this
geodesic to be unique). Now let $S$ be the distance between $0$ and $p$ along
the
geodesic, with it we define $\Gamma_{\alpha}=\partial
S/\partial\omega^{\alpha}$, which is then a vector, and $\Gamma^{a}=e^{\alpha
a}\Gamma_{\alpha}$. With those definitions, the external source $J^a$
transforms
like a vector and the whole eq.6 becomes covariant.

When the $\sqrt{g}$ factor in the measure is re-exponenciated it gives rise to
a new term in the lagrangian of the form:
\begin{equation}
\Delta{\cal L}=-\frac{i}{2}\delta^n(0)tr\; \log g
\end{equation}
The $\delta^n(0)$ can be
written as $\int d^nk/(2\pi)^n$, for this reason it is extremely
convenient to use dimensional regularization  when dealing with this kind of
models, since it is well known that in this scheme
 one formally takes the space-time dimension to be $n=4-\epsilon$ and the
integral vanishes. Then one can simply forget about the measure factor in eq.6.
Of course, we could choose to work in other regularization schemes, in that
case
the contribution of the $\delta^n(0)$ term in the lagrangian will cancel
other contributions  which are also absent
in the dimensional regularization scheme [16].

The main properties of the Green functions derived from the generating
functional in eq.6 are the following:

 a) Although the amplitudes depend in general on the coordinates defined on
$K$,
the $S$-matrix elements do not, provided the new coordinates are related to the
old ones as $\omega'^{\alpha}=\omega^{\alpha}+f^{\alpha}(\omega)$ with $f$
being
analytical and $f(0)=0$ i.e. the vacuum corresponds to $\omega^{\alpha}=0$ in
 any coordinate system. Then the predictions that can be made on physical
process with the model are independent of the coordinates chosen on $K$.

 b) The $G$ invariance of the (regularized) $W[J]$ functional gives rise to the
corresponding Ward-Slavnov-Taylor identities which in this case are called the
low-energy theorems since they can predict the very low-energy dynamics of the
GB.

 c) The counterterms needed to absorb the divergences are also $G$ invariant.
As far as all these terms are included in ${\cal L}$ with their corresponding
couplings $c_i$, the theory is renormalizable in this generalized sense. The
only difference with other much more conventional theories such as $QCD$ is
that
here we have an infinite number of coupling constants. However, following the
original philosophy of the chiral lagrangians, since one is interested in the
low-energy dynamics it is enough to compute the Green functions only to some
given power of the external momenta. In this case, only a finite number of
terms and couplings contribute, so that, at this level, the theory is
completely
predictive.

  Now the question arises on which is the proper choice of the groups $G$
and $H$ defining the quotient space $K=G/H$ for the description of the GB modes
of the symmetry breaking sector of the SM. The conditions that these groups
should satisfy are the following:

 a) As we need three GB to give mass to the three observed gauge bosons
$W^{\pm}$
and $Z$ we have $k= dim K=g-h=3$.

b) $G$ must contain the group $SU(2)_L\times U(1)_Y$ in order to be able to
couple the symmetry breaking sector to the electroweak gauge boson.

c) In order to ensure the experimental relation $\rho \simeq 1$
 we require the custodial symmetry to be present i.e. the subgroup $H$ must
contain the custodial group $SU(2)_{L+R}$. This automatically yields $\rho =1$
when the gauge fields are switched off and also implies that the photon will
remain massless since $U(1)_{em}$ is contained in $SU(2)_{L+R}$ and therefore
in
$H$.

 In principle one could think that the above conditions on the $G$ and $H$
groups are not very restrictive but we will show that there is only one
possible choice of these two groups compatible with them. This can be seen as
follows: The maximum number of isometries
 of some given $k$-dimensional space is $k(k+1)/2$ (see for instance [17]). In
our case this number must be
 larger or equal than the dimension of the isometry group $G$ i.e.:
\begin{equation} \frac{k(k+1)}{2}\ge g=k+h \end{equation} But condition a)
requires $k=3$ and condition c) means $h \ge 3$ so that $6\ge 3+h\ge 6$ which
implies $h=3$ and then $H=SU(2)_{L+R}$. In addition $g=6$ and taking into
account condition b) we arrive to the conclusion that $G=SU(2)_L\times SU(2)_R$
is the only possibility. Thus $K=G/H$ is just $S^3$ i.e. the three dimensional
sphere. In the following we will concentrate in these choices for $G$ and $H$
which is the only consistent with conditions a),b) and c) and therefore the
only relevant for the description of the symmetry breaking sector of the SM as
a NLSM.

Once we know which is the quotient space we have to work with, namely $S^3$,
one can ask which are the more suitable coordinates to parametrize it. In
principle, the formalism that we are going to use  is completely covariant and
can be used independently of the precise coordinate choice. However,
the most commonly used are the following:

1) Standard coordinates, where:
\begin{equation}
g_{\alpha\beta}=\delta_{\alpha\beta}+\frac{\omega^{\alpha}\omega^{\beta}}
{v^2-\omega^2}
\end{equation}
 being $v$ the radius of the sphere (this magnitude corresponds
to the vacuum expectation value of the Higgs fields in the MSM so we have
$v\simeq 250 GeV$). These coordinates are obtained just imposing on the
standard flat four dimensional space of Cartesian
 coordinates $(\omega^1,\omega^2,\omega^3,\sigma)$ the spherical constraint
$\omega^2+\sigma^2=v^2$.

2) Chiral coordinates where the fact that $SU(2)=S^3$ is used to ensemble the
GB
fields in a single $SU(2)$ matrix $U(x)=\exp (i
\omega^{\alpha}\sigma^{\alpha}/2v)$. In this case the metrics is given by:
\begin{equation}
 g_{\alpha\beta}=\frac{-v^2}{4}tr\frac{\partial
U}{\omega^{\alpha}} \frac{\partial U^{\dagger}}{\omega^{\beta}}
\end{equation}

3) Finally, sometimes it is useful to use geodesic coordinates where
$\omega^{\alpha}=\Gamma^{\alpha}$. In these coordinates
 the coupling of the external source to the GB fields in eq.6 becomes
especially simple and they are the most natural from the
geometrical point of view. In any case, it is important to remark that even if
one does not use geodesic coordinates we always have
$\omega^{\alpha}=\Gamma^{\alpha}+O(\omega^2)$

The use of one or another of the above coordinate systems can make some of the
computations much easier in different situations. Nevertheless, as
stressed before,  in further developments we will use a covariant formalism
which will be valid for any coordinate choice provided all them
are analytical and keep the classical vacuum at $\omega^{\alpha}=0$.

\section{Gauging the NLSM}

Once the NLSM describing the GB dynamics has been defined, the next step is
to switch on the gauge fields. This can be done as usual
by turning the derivatives of the NLSM into covariant
derivatives and adding  the pure Yang-Mills terms for the gauge fields.
We define the covariant derivative through those
$G$  Killing vectors which correspond to the gauge group $SU(2)_L\times
U(1)_Y$.
As discussed in the previous section, these Killing vectors are obtained
from the transformation equations for the GB fields. For the $SU(2)_L\times
U(1)_Y$
transformations we have:
\begin{eqnarray}
\delta_L\omega^{\alpha}(x)=l^{\alpha}_{\;a}\epsilon_L^a(x)  \\ \nonumber
\delta_Y\omega^{\alpha}(x)=y^{\alpha}\epsilon_Y(x)
\end{eqnarray}
where $l^{\alpha}_{\;a}$  with $a=1,2,3$ are the Killing vectors corresponding
to the $SU(2)_L$ group and $y^{\alpha}$ corresponds to
$U(1)_Y$. In fact, $y^{\alpha}$ is just the third killing vector of the
group  $SU(2)_R$. Now, the covariant derivative for the GB fields is defined
as;
\begin{equation}
D_{\mu}\omega^{\alpha}=\partial_{\mu}\omega^{\alpha}-gl^{\alpha}_{\;a}W_{\mu}^a-g'y^{\alpha}B_{\mu}
\end{equation} where $W_{\mu}^a$ and $B_{\mu}$ are the  $SU(2)_L$ and $U(1)_Y$
gauge fields and $g$ and $g'$ are the corresponding gauge couplings. Note that
the gauge fields are vectors both in the Minkowsky  space
 and in the $S^3$ sense. Thus  the lagrangian of the gauged NLSM can  be
written as;
\begin{eqnarray}
 {\cal L}_g    &=&{\cal L}_{YM}^L+{\cal L}_{YM}^Y  \\ \nonumber
&+&\frac{1}{2}g_{\alpha\beta}(\omega)D_{\mu}\omega^{\alpha}
D^{\mu}\omega^{\beta} \\ \nonumber
 &+& higher \; covariant \; derivative \;
terms
\end{eqnarray}
 where ${\cal L}_{YM}^L$ and ${\cal L}_{YM}^Y$ are the
Yang-Mills lagrangians for the
 $SU(2)_L$ and $U(1)_Y$ gauge fields.

The above lagrangian is invariant under the $SU(2)_L$ and $U(1)_Y$ gauge
transformations:
\begin{eqnarray}
\delta\omega^{\alpha}&=&l^{\alpha}_{\;a}\epsilon^a_L(x)+y^{\alpha}\epsilon_Y(x)
\\ \nonumber
 \delta W_{\mu}^a&=&
\frac{1}{g}\partial_{\mu}\epsilon^a_L(x)+\epsilon_{abc}
W_{\mu}^b \epsilon_{Lc}(x) \\ \nonumber \delta
B_{\mu}&=&\frac{1}{g'}\partial_{\mu}\epsilon_Y(x)
\end{eqnarray}
 provided the
killing vectors $l_a=l^{\alpha}_{\;a}\partial/\partial\omega^{\alpha}$ and $y=
y^{\alpha}\partial/\partial\omega^{\alpha}$ satisfy:
\begin{eqnarray} \{
l_a,l_b  \}&=&\epsilon_{abc}l_c  \\ \nonumber \{ l_a,y  \}&=&0   \\ \nonumber
\{ y,y  \}&=&0
\end{eqnarray}
 which are the so called closure relations. Here
$\{x,y\}$ should be understood as the Lie brackets of the Killing vectors $x$
and $y$ i.e.:
\begin{equation} \{  x, y  \}=\frac{\partial
y^{\alpha}}{\partial\omega^{\gamma}}x^{\gamma}  - \frac{\partial
x^{\alpha}}{\partial\omega^{\gamma}}y^{\gamma}
\end{equation}
The closure
relations are required to ensure that the $l_a$ and $y$ Killing vectors
 are a realization of the $SU(2)_L$ and $U(1)_Y$ symmetries on the coset
manifold $S^3$. Since we are using a covariant formalism all the time, we
only need to check these relations for some particular coordinate choice
on $S^3$. For example, for standard coordinates it is not
difficult to show:
 \begin{eqnarray}
l^{\alpha}_{\;a}=-\frac{1}{2}(\epsilon_{\alpha
a\gamma}\omega^{\gamma}-\delta_{\alpha  a}\sqrt{ v^2-\omega^2})    \\
\nonumber y^{\alpha}=-\frac{1}{2}(\epsilon_{\alpha
3\gamma}\omega^{\gamma}-\delta_{\alpha  3}\sqrt{ v^2-\omega^2})
\end{eqnarray}
With this expression for the $l_a$ and $y$ Killing vectors it is
straightforward to check the closure relations in eq.15 directly and this is
all we need to show the $SU(2)_L \times U(1)_Y$ gauge invariance of the
lagrangian in eq.13 for any coordinate choice on $S^3$. Similarly, it is
possible
to state the Jacobi identity as follows: Let
$\delta_1, \delta_2$ and $\delta_3$ be three arbitrary gauge transformations
like that in eq.15 with different parameters. Then we have:
\begin{equation}
[\delta_{\epsilon_1},[\delta_{\epsilon_2},\delta_{\epsilon_3}]]+
[\delta_{\epsilon_3},[\delta_{\epsilon_1},\delta_{\epsilon_2}]]+
[\delta_{\epsilon_2},[\delta_{\epsilon_3},\delta_{\epsilon_1}]]=0
\end{equation}
where the gauge transformation applies both to the GB and the gauge fields.

Thus, starting from the most general lagrangian for the NLSM (the one in eq.5)
it is possible to build an $SU(2)_L \times U(1)_Y$ gauged NLSM lagrangian
just replacing, as usual, the derivatives by covariant
derivatives and including the Yang-Mills terms for the gauge bosons. In
addition, we could consider other terms like:
\begin{equation}
F_{\mu\nu}^aF^{\mu\nu}_ag_{\alpha\beta}(\omega)D_{\rho}\omega^{\alpha}
D^{\rho}\omega^{\beta}
\end{equation}
 where $F_{\mu\nu}$  represents the standard stress tensor for the $SU(2)_L $
or
the $U(1)_Y$ gauge fields. In fact, some of these terms, which are gauge
invariant but not just obtained by the replacement of the derivatives by
covariant derivatives in the NLSM lagrangian, are needed in the quantized
theory
to cancel some of the divergences coming from the original (gauged) NLSM. In
any
case, it must be stressed that, even when one only considers the terms with two
covariant derivatives in eq.5, the $SU(2)_L \times U(1)_Y$ gauged NLSM is
no longer globally $SU(2)_L \times  SU(2)_R$  nor  $SU(2)_{L+R}$ invariant. In
particular, there are terms due to the $U(1)_Y$ gauge field which break these
symmetries and that is why the $\rho$ parameter gets electroweak
corrections that drive its value out of one when $g'$ is taken to be
different from zero, as it happens in the MSM.

	It is worth mentioning that
in the NLSM formulation of the standard model considered here it is not
necessary to go to unitary gauge to obtain the gauge bosons mass matrix
since for any coordinate choice on $S^3$ we have:
\begin{eqnarray}
l^{\alpha}_{\;a}&=&\frac{v}{2}\delta^{\alpha}_{a} +O(\omega)  \\  \nonumber
y^{\alpha}&=&-\frac{v}{2}\delta^{\alpha}_{3} +O(\omega)  \\  \nonumber
g_{\alpha\beta}&=&\delta_{\alpha\beta+}O(\omega ^2)
\end{eqnarray}
 and therefore, the
$g'^2g_{\alpha\beta}y^{\alpha} y^{\beta}B_{\mu}B^{\mu}/2$ term appearing in
eq.13 (inside the term with two covariant derivatives ) when expanded in powers
of the GB fields yields a $g'^2v^2B_{\mu}B^{\mu}/4$ piece contributing
to the mass matrix for any gauge and for any coordinate choice. However, we
still have terms like $gg_{\alpha\beta}\partial^{\mu}\omega^{\alpha} W_{\mu}^a
l_{\beta a}/2$  which contains in its expansion the
unwanted contribution
$g^2v\partial^{\mu}\omega^{\alpha}W_{\mu\alpha}/4$. Nevertheless, these mixing
terms can be cancelled in the quantized theory using the t' Hooft gauge
condition ($R_{\xi}$-gauges).

To end this section, we  will introduce some notations in order to save a lot
of space in next sections by considering the $SU(2)_L$ and $U(1)_Y$ gauge
groups
at the same time. To do that we introduce the Killing vector $L^{\alpha}_{\;a}$
with $a=1,2,3,4$ as $L^{\alpha}_{\;a}=gl^{\alpha}_{\;a}$ for $a=1,2,3$ and
$L^{\alpha}_{\;4}=g'y^{\alpha}$. We also introduce the completely antisymmetric
symbols $f_{abc}$ with $a=1,2,3,4$ as
 $f_{abc}=g\epsilon_{abc}$ for $a=1,2,3$ and $f_{ab4}=0$, and finally the gauge
field $W_{\mu}^a$ with $a=1,2,3,4$ will be defined as $W_{\mu}^a=W_{\mu}^a$ for
$a=1,2,3$ and $W_{\mu}^4=B_{\mu}$. With this notation the two covariant
derivative term in eq.13 can be written as:
\begin{eqnarray}
\frac{1}{2}g_{\alpha\beta}(\omega)D_{\mu}\omega^{\alpha}
D^{\mu}\omega^{\beta}= \hspace{8cm} \\ \nonumber
\frac{1}{2}g_{\alpha\beta}(\omega)\partial_{\mu}\omega^{\alpha}
\partial^{\mu}\omega^{\beta}
-g_{\alpha\beta}\partial_{\mu}\omega^{\alpha}L_{\alpha a}W^{\mu
a}+\frac{1}{2}g_{\alpha\beta} L^{\alpha}_{\;a}L^{\beta }_{\;b}W^{\mu a}W^{\mu
b}
\end{eqnarray}
 therefore, the mass matrix is just the zeroth
order term of $g_{\alpha\beta} L^{\alpha }_{\;a}L^{\beta }_{\;b}$ when expanded
in powers of the GB fields. The closure
 relations in eq.15 can now be gathered in:
\begin{equation}
\{ L_a,L_b  \}=f_{abc}L^c
\end{equation}

\section{The BRS symmetry and the quantized lagrangian}

The following step of our program is to build a
quantized version of our classical gauged NLSM appropriate for doing
perturbative computations. When a renormalizable gauge theory is described with
the fields belonging to linear representations of the gauge group
 and linear gauge fixing conditions, the Faddeev-Popov method using a standard
't Hooft gauge is very appropriate for this task providing a BRS invariant
lagrangian from which we could derive the Ward-Slavnov-Taylor identities that
will lead us to the ET.

 However, in order to make the proof as general as possible, we are interested
in
building a Lagrangian covariant under changes of coordinates in the GB coset.
As we will see, a crucial point in the proof of the ET is the usage of
a 't Hooft gauge fixing function. Usually, this function depends linearly
on the GB fields which are coordinates in the manifold that, in general,
do not transform like vectors in the coset index, and therefore will not
render a gauge fixing term  covariant under reparametrizations. Thus, if
we insist in demanding the lagrangian to be covariant under coordinate
changes in the coset, the 't Hooft gauge fixing function should depend
nonlinearly in the GB fields. It is known that Yang-Mills theories
quantized with non-linear gauge fixing conditions generate radiatively quartic
ghost interactions. That is why we  will use the technique described in [18]
that
generalizes the Fadeev-Popov method and introduces naturally this quartic
ghost terms through the BRS and anti-BRS invariance of the quantized
lagrangian. It is important to remark that if we had chosen to work with a
particular set of coordinates, the GB fields could have appeared linearly in
the 't Hooft gauge fixing function, and then we could have used the standard
Fadeev-Popov procedure, but that would not imply that the results could be
translated directly to other coset coordinates.

 The quantization method goes as follows: We start from a gauge invariant
lagrangian like that considered in the previous section where the fields do not
necessarily transform linearly. Then, by introducing anticommuting ghost fields
one obtains the corresponding BRS and anti-BRS  transformation (at this point
the
closure and Jacobi identities in eq.15, eq.18 and eq.22 play an decisive
role). Next we build the so called quantum lagrangian, BRS and anti-BRS
invariant, which is a generalization of the standard classical lagrangian plus
the usual gauge fixing and Faddeev-Popov terms. Therefore, the Feynman rules
and a well defined perturbative expansion for the gauged NLSM can be derived as
it is usually done.

Thus we start introducing the ghost fields $c^a$ and $\bar c^a$ where the
flat index $a=1,2,3$ refers to the $SU(2)_L$ ghost fields and $a=4$ refers
 to that from $U(1)_Y$. They are scalar fields and anticommuting
Grassmann variables. Now, the BRS transformations for the GB, gauge bosons and
the ghost fields can be written as:
\begin{eqnarray}
s[\omega^{\alpha}]&=&L^{\alpha}_{\;a}c^a   \\  \nonumber
s[W^{\mu a}]&=&\partial^{\mu}c^a+f^a_{\;bc}W^{\mu b}c^c\equiv D^{\mu}_{\;ac}c^c
\\  \nonumber
s[c^a]&=&-\frac{1}{2}f^a_{\;bc}c^bc^c
\end{eqnarray}
 and the anti-BRS  ones as:
\begin{eqnarray}
\bar s[\omega^{\alpha}]&=&L^{\alpha}_{\;a}\bar c^a   \\  \nonumber
\bar s[W^{\mu a}]&=& D^{\mu}_{\;ac}\bar c^c \\  \nonumber
\bar s[\bar c^a]&=&-\frac{1}{2}f^a_{\;bc}\bar c^b \bar c^c
\end{eqnarray}
 and in addition, we define the combined relation;
\begin{equation}
 s[\bar c^a]+\bar
s[ c^a]=-f^a_{\;bc}\bar c^b \bar c^c
\end{equation}
Still we have to define the
BRS transformation for the antighost field and the anti-BRS  for the ghost
field. To do that we introduce the auxiliary commuting scalar field $b^a$.
Then we write:
\begin{eqnarray}
s[\bar c^a]&=&b^a   \\ \nonumber s[b^a]&=&0
\end{eqnarray}
 and
\begin{eqnarray}
\bar s[\bar c^a]&=&-b^a-f^a_{\;bc}\bar c^b c^c   \\ \nonumber
\bar s[b^a]&=&-f^a_{\;bc}\bar c^b
c^c
\end{eqnarray}
This completes the set of BRS
and anti-BRS  transformations. The action of the BRS and the anti-BRS
 operators $s$ and $\bar s$ on more complex combinations of fields can be
defined assuming
 that these are linear differential operators graded by the ghost number so
that we have $s[XY]=s[X]Y\pm Xs[Y]$ where the minus sign applies if
there is an odd number of ghost or anti ghost  in the product of fields $X$
(note that the auxiliary field $b^a$ has zero ghost number).

With the above definition for the BRS and the anti-BRS  operators it is
possible
to show that they are nilpotent in the following sense
\begin{equation}
 s^2=s \bar s+\bar s s=\bar s ^2=0
\end{equation}
 In  order to  show these nilpotency
relations  we have to verify that they hold for all the fields. The
computations are tedious but straightforward provided one uses eq.23 eq.24,
eq.25, eq.26 and eq.27, the other properties of the $s$ and the $\bar s$
operators and the closure and Jacobi identities in eq.15 or eq.22  but the
details
will not be given here. As the (anti-) BRS transformation properties for the
physical fields (the GB and the gauge bosons) can be understood as gauge
transformations where the gauge parameter is just the (anti-) ghost field, one
could say that the original gauge transformations plus the closure and the
Jacobi identities are equivalent to the (anti-) BRS transformation plus the
nilpotency relations realized in the enlarged system formed by the physical
fields, the (anti-) ghosts and the auxiliary field. However, as it is well
known, the (anti-) BRS version of the gauge transformations is by far much more
appropriate for the quantum theory, and in particular for perturbation theory.
As a matter of fact, the physical Hilbert space can be defined as the
cohomology
of the BRS operator, i.e. as the set of states which are annihilated by the BRS
operator not being the result of the action of the BRS operator on any other
state. In addition, the complete antisymmetry  of the $f^a_{\;bc}$ symbols
ensures the existence of an invariant functional integral measure for the
fields
of the enlarged system.

Now we will define the quantum lagrangian as the most
general one being covariant (in the space time and the $S^3$ sense), $s$ and
 $\bar s$ invariant and with ghost number zero . It can be written as:
 \begin{equation}
{\cal L}_Q={\cal
L}_g+\frac{1}{2}s \bar s[G(\omega^{\alpha},W^a_{\mu},c^a,\bar c^a,b^a)]
\end{equation}
 where $G$ is some scalar function of all the fields with zero ghost number
 and ${\cal L}_g$ is the classical gauge invariant lagrangian (the factor $1/2$
is included for further convenience). The (anti-) BRS invariance of the above
quantum lagrangian follows from the gauge form of
these transformations for the physical fields and the nilpotency relations of
the (anti-) BRS operator.  Different choices of $G$ correspond to different
gauge fixing functions for the quantized theory. In our case, we are dealing
with a gauge theory which is spontaneously broken and for this reason the t'
Hooft or renormalizable
 gauge fixing conditions ($R_{\xi}$-gauges) are very appropriate. These kind of
gauges     have the two following
 essential properties; first they give rise in the quantum lagrangian to a
bilinear term for the gauge fields which is invertible so that we have a well
defined gauge field propagator that can be used in perturbation theory. Second,
a piece appearing in the quantum lagrangian cancels the unwanted mixing term
which appears in the classical lagrangian (inside the second term in the RHS of
eq.21). In the framework of the BRS formalism considered here, the
$R_{\xi}$-gauges can be obtained using the function:
\begin{equation}
G=AW_{\mu}^aW^{\mu a}+Bf(\omega)+Cc^a\bar c_a
\end{equation}
where $f$ is some
arbitrary scalar function on $S^3$ i.e. on the GB fields with;
\begin{equation}
\frac{\partial f}{\partial\omega^{\alpha}}=\Gamma_{\alpha}+O(\omega^2)
\end{equation}
The constants $A,B$ and $C$ will be adjusted to cancel the unwanted mixing
term coming from the classical lagrangian. Giving the appropriate values to
these constants we obtain:
\begin{equation}
{\cal L}_Q={\cal
L}_g+\frac{1}{2}s \bar s[W_{\mu}^aW^{\mu a}-2\xi f(\omega)+\xi c^a\bar c_a]
\end{equation}
 Now, working out the (anti-) BRS transformation we finally
obtain the following quantum lagrangian;
\begin{eqnarray}
{\cal L}_Q &=&{\cal L}_g- \frac{1}{2\xi}[\partial_{\mu}W^{\mu
a}+\xi\frac{\partial f}{\partial\omega^{\alpha}} L^{\alpha a}]^2 \\ \nonumber
&+& D_{\mu b}^a c^b\partial^{\mu}\bar c_a -\xi
\{\frac{\partial^2f}{\partial\omega^{\alpha}\partial\omega^{\beta}}
L^{\alpha}_{\;a}L^{\beta}_{\;b} + \frac{\partial f}{\partial\omega^{\alpha}}
\frac{\partial L^{\alpha}_{\;a}}{\partial\omega^{\beta}} L^{\beta}_{\;b} \}
c^b \bar c^a        \\ \nonumber
&+& \frac{1}{2}[\bar c,c]_a(\partial_{\mu}W^{\mu a}+\xi\frac{\partial f}
{\partial\omega^{\alpha}}L^{\alpha a}) + \frac{\xi}{4}[\bar c, c]^2
\end{eqnarray}
where $c=T^a c^a$ and $\bar c = T^a \bar c^a$ being $T^a=\sigma
^a/2$ for $a=1,2,3$ and $T^4=1/2$
 and $[c, \bar c]=[c, \bar c]^aT^a$. The gauge fixing function has appeared in
the second term of the RHS of the above lagrangian since the $b$ field
equation of motion is given by:
\begin{equation}
b^a=\partial_{\mu}W^{\mu
a}+\xi\frac{\partial f}{\partial\omega^{\alpha}}L^{\alpha a}
\end{equation}
 so that the auxiliary
$b^a$ plays the role of a Lagrange multiplier needed  to implement the gauge
fixing condition. The gauge fixing terms give rise to the standard
$R_{\xi}$
 propagator for the gauge fields and it is not difficult to check that it
cancels the unwanted terms in the classical lagrangian using eq.31 and
remembering that
 $\omega^{\alpha}=\Gamma^{\alpha}+O(\omega^2)$ for any coordinate system. The
following four terms
 are the generalization of the standard Faddeev-Popov method including the
kinetic part for the
 (anti-) ghost fields and their interactions with the gauge  and the GB.
 Finally, we have the last
 contribution which produces quartic ghost interactions. This term cannot be
obtained by the standard Faddeev-Popov procedure. However, it is radiatively
generated in Yang-Mills theories quantized
 with non-linear gauge fixing functions. In our case, it comes from the
$c\bar c$ piece in our $G$ function through a $b^2$ term (the
gauge fixing condition), being a genuine element in the formalism used here.

 In order to be sure of our results after applying the $s$ and  the $\bar s$
operator to our $G$ function we have also checked directly that the whole
quantum
lagrangian in eq.33 is (anti-) BRS invariant as it obviously should be.

Thus we have succeeded in the construction of a quantum lagrangian
corresponding to the
the most general gauged NLSM describing the dynamics of the GB and the gauge
fields. The Feynman rules to be used in perturbation theory can be derived as
usual from this
 quantum lagrangian. Of course, the precise form of these rules depends
on the coordinates one chooses in the coset $S^3$. However, our quantum
lagrangian is completely
covariant and (anti-) BRS invariant and it is independent of the coordinates
and
the gauge provided it is renormalizable. At this point it is worth noting that
for the particular choice of the Landau gauge, which corresponds to take
$\xi=0$, the quantum lagrangian in eq.33 becomes greatly simplified and, in
particular, there are
 no interactions between the (anti-) ghost and the GB fields.

\section{The renormalized lagrangian}

Now we can derive the Ward-Slavnov-Taylor identities for the dimensionally
regularized Green functions that are needed in the proof of the ET, making use
of the BRS invariance of the lagrangian in eq.33. It is important to remark
that
dimensional regularization is used in order to avoid the
$-\frac{i}{2}\delta^n(0)tr\; \log g$ contribution from the path integral
measure
of the GB fields, as well as to preserve the (anti)-BRS
invariance in the regularized lagrangian.

	However, for practical
purposes we are interested in the renormalized Green functions in order to make
predictions for the different physical processes. The Green functions obtained
from the lagrangian in eq.33 present divergencies that have to be cancelled by
considering a renormalized lagrangian made of the original "bare" lagrangian of
eq.33 plus the counterterms needed to reproduce the whole set of divergent
structures. Although these counterterms have been given only up to those with
four derivatives [19], all them should also be (anti)-BRS invariant, since if
this was not the case the model would be anomalous, i.e., the gauge invariance
of the model would be broken by quantum effects.  However, the standard
hypercharge assignments in the SM, and the fact that the number of colors is
$N_c=3$, ensures the cancellation, generation by generation,  of all  the
possible gauge and  mixed gauge-gravitational anomalies, including the
non-perturbative $SU(2)$ discovered by Witten [20]. Therefore, even though we
coupled chiral fermions to the NLSM the resulting theory would be free of these
anomalies. Apart from the (anti)-BRS invariance, the model we have built is
also
invariant under reparametrizations of the GB, since it has been shown in [21]
that the potential anomalies that could break the invariance under changes of
coordinates on the coset, are absent from the NLSM when it is defined in
spaces
with a  dimension smaller than that of the space-time.

Therefore, although the (anti)-BRS invariant lagrangian obtained  when taking
into account all the needed counterterms is made of an infinite number of
terms, it can be understood as the renormalized lagrangian of a renormalizable
theory with an infinite number of couplings written in terms of bare
quantities.
As it happens in any renormalizable theory, we can also write this renormalized
lagrangian by means of  renormalized fields and
couplings, so that the terms which appear
now will keep the same form (due to the fact that the theory is renormalizable
in
the sense described above) although they present some $Z$ factors. The relation
between bare and renormalized fields and gauge couplings is given by:

\begin{eqnarray}
W_{0\mu}^a(x) =Z_3^{(a)1/2}W_{\mu}^a(x) ;
\omega_o^{\alpha}(x)=Z_{\omega}^{(\alpha)1/2}\omega^{\alpha}(x) ;
g_0^{(a)}=Z_g^{(a)}g^{(a)} ;
\xi_0^{(a)} =Z_3^{(a)}\xi^{(a)}           \\   \nonumber
c_0^a(x) =\widetilde Z_2^{(a)1/2}c^a(x) ;
\bar c_0^a(x)=\widetilde Z_2^{(a)1/2} \bar c^a(x) ;
B_0^a(x)=\widetilde Z_2^{(a)}B^a(x)
; v_0 =Z_v^{1/2}v
\end{eqnarray}

where $g^{(a)}=g$ for $a=1,2,3$ and $g^{(4)}=g'$. The first three $Z_3$ are
the same thanks to the gauge structure of the model. As a
notation, in the following we will
 use that the indices between parenthesis are not summed
and that those  fields and constants without $0$ subscripts are renormalized.
We
have also introduced for further convenience the $B$ field which is the $b$
auxiliary field with a different normalization: $B_0^a=\sqrt{\xi_0}b_0^a$. In
addition, the symmetries of the chiral lagrangian provide infinite relations
between the bare and the renormalized couplings. Indeed, thanks to the
(anti)-BRS invariance of the renormalized lagrangian given in terms of the bare
quantities, it is possible to find the following "renormalized" (anti)-BRS
transformations which will leave invariant the renormalized lagrangian once
it is written in terms of the renormalized fields and couplings:
\begin{eqnarray}
s_R[\omega^{\alpha}]=X^{(a)}L^{\alpha}_{Ra}c^a
\;\;\;\;\;\; &\qquad&   \bar s_R[\omega^{\alpha}]=X^{(a)}L^{\alpha}_{Ra}\bar
c^a
\\  \nonumber  s_R[W^{\mu a}]=X^{(a)}D^{\mu a}_{Rc}c^c \;\;\; &\qquad& \bar
s_R[W^{\mu a}]=X^{(a)}D^{\mu a}_{Rc}\bar c^c \\  \nonumber
s_R[c^a]=-\frac{X^{(a)}}{2}f^a_{R\;bc}c^bc^c &\qquad&  \bar
s_R[c^a]=-\frac{X^{(a)}}{2}f^a_{R\;bc}\bar c^bc^c \\ \nonumber s_R[\bar
c^a]=X^{(a)}\frac{B^a}{\sqrt{\xi^{(a)}}} \;\;\;\;\;\;\; &\qquad& \bar
s_R[\bar c^a]=-X^{(a)} \left( \frac{B^a}{\sqrt{\xi^{(a)}}} +
f^a_{R\;bc}\bar c^bc^c \right) \\ \nonumber  s_R[ B^a]=0
\;\;\;\;\;\;\;\;\;\;\;\;\;\;\;\;\;\;\; &\qquad& \bar s_R[ B^a]=-X^{(a)}
f^a_{R\;bc}\bar c^bB^c \\  \nonumber \end{eqnarray}

 where $L^{\alpha}_{Ra}=Z_{\omega}^{(\alpha)-1/2}Z_3^{(a)1/2}L^{\alpha}_{\;a}$,
 $f^a_{R\;bc}=Z^{(a)}_g Z_3^{(a)1/2} g f^a_{\;bc}$, and $X^{(a)}=\widetilde
Z_2^{(a)1/2} / Z_3^{(a)}$.

 Once we have a set of (anti)-BRS
 symmetry transformations for the renormalized fields, and a renormalized
lagrangian which is invariant under such transformations, we can obtain
Ward-Slavnov-Taylor identities for the renormalized Green functions by means of
the standard functional methods. In particular, we are going to use some of
these relations to find the actual formulation of the ET that
 holds when the symmetry breaking
 sector of the SM is described through the chiral lagrangian formalism.

\section{Ward-Slavnov-Taylor Identities}

 As discussed in the introduction, our aim in this section is to use
the general functional procedures to obtain Ward-Slavnov-Taylor identities
 showing the relationship between renormalized Green functions involving an
arbitrary number of longitudinal gauge bosons $W_L$, with those
Green functions where
 we have replaced all the external $W_L$ by their corresponding GB. In order
to do that  we will basically follow the same steps of the Chanowitz-Gaillard
proof of the ET.

In the preceding section, we have explicitly
 built an (anti-) BRS invariant renormalized quantum lagrangian for the
$SU(2)_L\times U(1)_Y$ gauged NLSM. Now we can use it to obtain
Ward-Slavnov-Taylor identities involving Green functions which will lead us to
the desired relations for $S$ matrix elements. To do so, we start from the
standard identity derived from the lagrangian BRS invariance (here $A$ stands
for any of the renormalized fields appearing in the quantum lagrangian i.e.
$A_i=\omega^{\alpha}, W^a_{\mu}, c^a, \bar c^a, B^a$):
\begin{equation}
\sum_i \int d^4x <s_R[A_i]>_J J_i(x)=0
\end{equation}
Where the BRS transformations are made of linear and nonlinear products of
fields, so that, in general, we can write :
\begin{equation}
<s_R[A_i]>_J=\sum_n s_{A_i}^{i_1...i_n} <A_{i_1}...A_{i_n}>_J
\end{equation}
In most of the cases we only have to take into account the linear terms,
but when $A_i=W^{\mu},c^a$ we have to consider products of two fields,
and when $A_i=\omega^{\alpha}$ all the infinite terms that appear due to the
nonlinear realization of the symmetry.
 Now we can relate these expectation values with the generating functional
for connected Green functions $W_R(x_1,...,x_n)$, using the standard functional
 formalism:
\begin{equation}
<A_{i_1}...A_{i_n}>_J=\frac{\delta^{(n)}W_R[J]}{\delta J_{i_1}...\delta
J_{i_n}}
\end{equation}
where, as usual, $J_i$ is the external current associated with the $A_i$ field,
and the generating functional is given by:
\begin{eqnarray}
 W_R[J]&=&\sum_{n=1} \int
d^4x_1d^4x_2...d^4x_n W_{R \; i_1,...,i_n}(x_1,x_2,...,x_n)
J_{i_1}(x_1)...J_{i_n}(x_n) \\  \nonumber &=& (2\pi)^4 \sum_{n=1} \int
\prod_{i=1}^n \frac{d^4p_i}{(2\pi)^4} \delta^4(\sum_i
p_i)J_{i_1}(-p_1)...J_{i_n}(-p_n) W_{R \; i_1,...,i_n}(p_1,...p_n)
\end{eqnarray}
(from now on we will be working in momentum space).
 Using the preceding relations, the condition of
BRS invariance in eq.37 becomes:
\begin{equation}
I[J] = \sum_i \sum_n s_{A_i}^{i_1...i_n} \int
\frac{d^4q d^4k_1...d^4k_{n-1}}{(2\pi)^{4n}}
 \frac{\delta^{(n)}W_R[J]}{\delta J_{i_1}(q-k_1)...\delta J_{i_n}(k_{n-1})}
 J_i(-q) =0
\end{equation}

A similar equation can be written for the renormalized anti-BRS
transformations too. It is
straightforward then to obtain Ward-Slavnov-Taylor identities from the last
formula just by taking functional derivatives with respect to $J_i(p)$ and
setting then $J=0$. In particular, since we want to relate external $W_L$ with
GB, we are interested in identities involving the auxiliary $B$ field because,
as we have seen yet, when using its equation of motion it gives the gauge
fixing
condition which
 is hinting us the possible relation between longitudinal vector bosons and
GB. If we look again to the BRS transformations, we find that
the $B$ field only appears in $s_R[\bar c]$, and as we are looking for a
relation
where all the $W_L$ are replaced by $\omega$, we need to derive functionally
once by $\bar c$.

As an example, and since we will need the result for the general proof, we are
going to obtain a relation with just one $B$ and another arbitrary field.
Besides, it illustrates some differences with the proof given by
Chanowitz-Gaillard. We start from:
\begin{equation}
 \left.\frac{\delta}{\delta
J_{\bar c_b}(-k)}\frac{\delta}{\delta J_j(p)} I[J] \right|_{J=0} =0
\end{equation}
 Noticing that the only possible contributions come from $A_j=W^{\mu}$
or $A_j=\omega^{\alpha}$, we obtain the following relations:
\begin{eqnarray}
 X^{(a)}( L_{Ra}^{\alpha (0)} + \Delta_{2a}^{\alpha}(p^2))
W_{c^{a}\bar c^{b}}(p)
+ \frac{X^{(b)}}{\sqrt {\xi^{(b)}}} W_{B^{b}\omega^{\alpha}}(p)=0 \\ \nonumber
 X^{(a)} ip_\mu(1+\Delta_3(p^2)) W_{c^{a}\bar c^{b}}(p) +
\frac{X^{(b)}}{\sqrt {\xi^{(b)}}} W_{B^{b}W^a_{\mu}}(p)=0
\end{eqnarray}
Where we have expanded $L_{Ra}^{\alpha}$ as:
\begin{equation}
L_{Ra}^{\alpha}=L_{Ra}^{\alpha (0)}+L_{Ra}^{\alpha \beta (1)}
\omega ^{\beta}+L_{Ra}^{\alpha \beta \gamma(2)} \omega^{\beta} \omega^{\gamma}
+...
\end{equation}
and we have used the following definitions:
\begin{eqnarray}
ip_{\mu}\Delta_3(p^2)&=&f^a_{Rdc}W^{-1}_{c^{a}\bar c^{b}}(p) \int
\frac{d^4}{(2\pi)^4}
W_{W^{\mu d}c^{c}\bar c^{b}}(p-q,q,p)\\ \nonumber
\Delta_{2a}^{\alpha}(p^2)&=& L_{Ra}^{(1)\alpha \beta}W^{-1}_{c^{c}\bar
c^{b}}(p) \int \frac{d^4}{(2\pi)^4}
W_{\omega^{\beta}c^{c}\bar c^{b}}(p-q,q,p) + ...
\end{eqnarray}
	Therefore we can gather these results in:
\begin{equation}
 \frac{X^{(b)}}{\sqrt{\xi^{(b)}}}W_{B^{b}l}(p)
=- X ^{(a)}D^{a}_{ R l}(p) W_{c^{a}\bar c^{b}}(p)
 \end{equation}
where:
\begin{equation}
D^{a}_{ R l}(p) = ip_{\mu}(1+\Delta_3(p^2))\delta_l^{W_{\mu a}}+
(L^{(0) \alpha}_{R a}+ \Delta_{2a}^{\alpha}(p^2))
\delta_l^{\omega^{\alpha}}
\end{equation}
 and in the last step we have introduced the ${D}^{a}_{Rl}(p)$
operator which will play a similar role to the $D^{a}_l(p)$ operator defined by
Chanowitz and Gaillard, although we can see yet some important differences with
their formal proof: First, the $L_R^{(0)}$ factor coming from the nonlinear
realization of the symmetry, which appears only in its zeroth order and in the
end will be the only remainder of the complicated relation between gauge and
Goldstone bosons
that one naively expects from the nonlinear gauge fixing condition. Second,
the $X$ factors due to renormalization. And third, the $\Delta$ terms that
were correctly introduced by Bagger and Schmidt [14] in the context of the
MSM.\footnote{ Recently we have noticed [22] a paper where these $\Delta$ terms
were correctly introduced in the chiral lagrangian formalism}.

 Now, in order to derive the general expression we start from:
\begin{equation}
 \left.\frac{\delta}{\delta J_{\bar c_{a_1}}(-k)}
\prod_{j=2}^{s}\frac{\delta}{\delta J_{B_{a_j}}(-k_j)}
\prod_{k=1}^{m}\frac{\delta}{\delta J_{A_k}(-p_k)}
I[J] \right|_{J=0} =0
\end{equation}
 Where we will impose to the $J_{A_k}$ currents  to be associated to physical
$A_k$ fields only. Therefore, as we are not taking functional derivatives with
respect to $J_{\omega}$ nor $J_c$, we can easily see from the BRS
transformations that we do not get any contribution when $A_i=B$, neither when
$A_i=\omega,c$. Since the $A_k$ are physical, their polarization vectors
satisfy
$\epsilon \cdot k_\mu =0$ and thus they will cancel the first term coming from
$s_R[W^a_{\mu}] = ik_\mu c^{a}+\epsilon^a_{Rbc}W_{\mu b}c_c$. Therefore we only
have to take into account those contributions coming from $s_R[\bar c]$ and the
part which is left from $s_R [W^a_{\mu}]$ that will be called  generically,
"bilinear terms".  Finally we can write:
\begin{equation}
 \frac{X^{(a_1)}}{\sqrt {\xi^{(a_1)}}}
W_{B_{a_1}B_{a_2}...B_{a_s}A_1...A_m}(k_1,...,k_s,p_1,...p_m) + \mbox{
bilinear terms} =0
\end{equation}
where $\sum_i k_i =-\sum_i p_i$. In order to translate this result to off-shell
$S$-matrix elements, we apply the  Lehmann-Symanzik-Zimmermann (LSZ) reduction
formula which yields:
\begin{eqnarray}
  \frac{X^{(a_1)}}{\sqrt {\xi^{(a_1)}}} \left( \prod_{i=1}^{m}W_{A_iA_i}
(p_i) \right) \sum_{l_j} \left( \prod_{j=1}^{s}W_{B_{a_j}l_j}(k_j) \right)
S^{off-shell}_{l_1..l_s A_1...A_m}(k_1...k_s,p_1...p_m)  \\ \nonumber
 + \mbox{ bilinear terms}=0
\end{eqnarray}

	Note that the $a_1$ index is not contracted so that the factor
 $X/\sqrt{\xi}$ is not relevant and we can drop it. The next step to obtain
$S$-matrix elements is to multiply by the inverse renormalized propagators
of the $A_i$ fields the whole eq.50. When we set their momenta on shell, that
is
$p^2_i=m_{A_i}^2$, we can cancel the "bilinear terms" since they
contain a Green function made of more than $s+m$ fields, so that
 one of the momenta is off-shell, and therefore the pole needed to
compensate for $W_{A_iA_i}^{-1}(p_1) \rightarrow 0$ is absent and the
term vanishes as $p_1^2\rightarrow m_{A_1}^2$. Finally, we can use eq.46 to
substitute the $B$ field two point functions, and then write:

\begin{equation}
\left. \sum_{l_j} \left(\prod_{j=1}^{s}\frac{\sqrt {\xi^{(a_j)}}}{X^{(a_j)}}
X^{(c_j)} W_{c^{c_j}\bar c^{a_j}}(k_j) D^{c_j}_{ R l_j}(k_j) \right)
S^{off-shell}_{l_1..l_s A_1...A_m}(k_1...k_s,p_1...p_m)
\right|_{p^2_i=m_{A_i}^2} =0
\end{equation}

 Again, the $a_j$ are not contracted, and therefore, we can take away the
$\sqrt
{\xi^{(a_j)}}/X^{a_j}$ factors.  Now, although the renormalized ghost
two point function may be in principle non-diagonal, we can
multiply by its inverse $W^{-1}_{c^{d_j}\bar c^{a_j}}(k_j)$ so that the $d_j$
index is set free thus allowing us to drop the last $X$ factor. Therefore, we
arrive to the following expression:
\begin{equation}
\left. \sum_{l_1...l_r}\prod_{i=1}^{s}D^{a_i}_{R l_i}(p_i) S^{off-shell}
_{l_1..l_s,A_1..A_m}(p_1..p_s,k_1..k_m) \right|_{k^2_i=m_{A_i}^2} =0
\end{equation}

\section{The Equivalence Theorem}

Our aim is now to complete the LSZ procedure to obtain from eq.52 relations
between $S$-matrix elements by setting all the momenta
on-shell. However, before doing so, we have to obtain the physical
field combinations since they are not the $W_{\mu}$ fields in the $D_R$
operator.
This step is achieved by means of a transformation  $\widetilde W^a_{\mu}=
R^{ab}W^b_{\mu}$, which can be given in the most general form by:
\begin{equation}
\left ( {\matrix{ \widetilde W_{\mu}^1 \cr  \widetilde
W_{\mu}^2 \cr \widetilde W_{\mu}^3 \cr
\widetilde W_{\mu}^4  }} \right ) =
\left ( {\matrix{ W_{\mu}^- \cr W_{\mu}^+ \cr Z^{phys}_{\mu} \cr A^{phys}_{\mu}
}}
\right) = \left ( {\matrix{ 1/\sqrt{2}&i/\sqrt{2}&0&0\cr
1/\sqrt{2}&-i/\sqrt{2}&0&0\cr
0&0&cos \theta & -sin \theta \cr
0&0&sin \theta' & cos \theta'  }} \right )
\left ( {\matrix{ W_{\mu}^1 \cr W_{\mu}^2 \cr W_{\mu}^3 \cr W_{\mu}^4
 }} \right)
\end{equation}
The renormalized fields thus obtained are precisely those which ensure that
their exact propagators have poles located at the right values of their
corresponding physical masses. Similarly we also define:
$\widetilde L^{(0)b}_{R\alpha}=L^{(0)a}_{R\alpha} (R^{-1})^{ba}$,as well as
$\widetilde \Delta_{2a}^{\alpha}$, and so we write:

\begin{equation}
\sum_{l_1...l_r}\prod_{i=1}^{s}\widetilde{D}^{a_i}_{R l_i}(p_i)
S _{l_1..l_s,A_1..A_m}(p_1..p_s,k_1..k_m) = 0
\end{equation}
where
\begin{equation}
\widetilde{D}^{a}_{ R l}(p) = ip_{\mu}(1+\widetilde \Delta_3(M_{phys}^2))
\delta_l^{\widetilde{W}_{R \mu a}}+
(\widetilde{L}^{(0) \alpha}_{ Ra} + \widetilde
\Delta_{2a}^{\alpha}(M_{phys}^2))
\delta_l^{\omega^{\alpha}}
\end{equation}
Note that we have already set the $p_i$ momenta on-shell for the
massive physical vector bosons i.e. $p_i^2=M_{i \;\; phys}^2$. Thus this will
also be  the on-shell conditions for the GB in the final version of the ET.

	The next step is now to convert the momentum factors of
$\widetilde{D}^{a}_{Rl}(p)$ in longitudinal polarization vectors
$\epsilon_{(L)}=p_\mu / M_{phys}+v_\mu$ since we expect to neglect the $v_\mu$
four-vector at sufficiently high energies because $v_\mu \simeq O(M_{phys}/E)$.
Unfortunately, this step is not straightforward. The gauge structure of the
theory will produce cancellations in the amplitudes involving longitudinally
polarized gauge bosons, and we are not allowed to simply drop out the terms
containing $v_{\mu}$ vectors.

The way around this problem is to perform a power counting analysis to
extract the relevant orders for the ET statement. It will be very convenient to
derive from eq.54 an expression involving only longitudinal polarization
vectors and $v_{\mu}$ factors, but without any momentum multiplying the
amplitudes. Such an expression was first obtained in [10] and, although we have
introduced the $K_R$, its derivation is completely analogous.
For the sake of completeness, we will sketch it here. We start defining:
\begin{equation}
V(l,n,m)= \left( \prod_{i=1}^{n}v_{\mu_i} \right)
\left( \prod_{j=1}^{l} K_{\alpha_j}^{a_j} \right)
S_{\omega^{\alpha_1}...\omega^{\alpha_l}; \widetilde{W}^{b_1}_{R \mu_1}...
\widetilde{W}^{b_n}_{R \mu_n};A_1...A_m}(p;q;k)
\end{equation}
where we have introduced $K^{\alpha}_{Ra}=( \widetilde{L}^{(0)
\alpha}_{Ra} + \widetilde \Delta_{2a}^{\alpha}(M_{phys}^{(a)2})) /
(M^{(a)}_{phys}(1+\widetilde \Delta_3(M_{phys}^{(a)2})))$, which reflects
the effects of the
renormalization procedure and of the non-linear realization of the theory, and
thus, up to here, are the main difference with the ET proof given in [10].
It will be important to remark that these factors do not depend on the energy.
\footnote {When we
were completing this work we noticed [22] a paper where the renormalization
factors have been considered in the ET for the chiral lagrangian formalism}

If we write $v_\mu=\epsilon_{(L)}-p_\mu / M_{phys}$ we will obtain a sum whose
generic term with $s$ longitudinal polarization vectors and $n-s$ momentum
factors will be of the general form:
\begin{eqnarray}
X(l,n-s,s,m)=
\left( \prod_{i=1}^{n-s} \frac{q_{\mu_i}}{M_{b_i}} \right)
\left( \prod_{k=1}^{s} \epsilon_{(L)}^{\nu_k}(r_k) \right) \times
\hspace {5cm} \\ \nonumber
\left( \prod_{j=1}^{l} K_{\alpha_j}^{a_j} \right)
S_{\omega^{\alpha_1}...\omega^{\alpha_l}; \widetilde{W}^{b_1}_{R \mu_1}...
\widetilde{W}^{b_{n-s}}_{R\mu_{n-s}};\widetilde{W}^{c_1}_{R \nu_1}...
\widetilde{W}^{c_s}_{R\nu_s}A_1...A_m}(p;q;r;k)
\end{eqnarray}

	Indeed, when writing explicitly $V(l,n,m)$, these $X(l,n-s,s,m)$ terms will
appear with different contractions of indices, so that we can write:
\begin{equation}
\bar V(l,n,m) = \sum_{s=0}^{n} (-1)^{n-s}\bar X(l,n-s,s,m)
\end{equation}
 where the bar means a sum over all independent permutations of the indices
$(a_i,p_i)$, $(b_i,q_i)$, and $(c_k,r_k)$.

	With these notations it is easy to see that eq.54 can be written as:
\begin{equation}
\sum_{l=0}^{n-s} (-i)^{n-s-l} \bar X(l,n-s-l,s,m) = 0
\end{equation}
We can multiply this last equation by $i^s$, adding the result from $s=0$ to
$s=n-1$. Thus we obtain:
\begin{eqnarray}
0&=&\sum_{s=0}^{n-1} i^s \sum_{l=0}^{n-s} (-i)^{n-s-l} \bar X(l,n-s-l,s,m) \\
\nonumber
&=& \sum_{l=0}^{n} i^{n-l} \sum_{s=0}^{n-l} (-1)^{n-s-l} \bar X(l,n-s-l,s,m)
-i^n \bar X(0,0,n,m)
\end{eqnarray}
There is only one permutation in $\bar X(0,0,n,m)$ so that we can drop the bar,
and using eq.58 we finally arrive to:
\begin{equation}
X(0,0,n,m) = \sum_{l=0}^{n} (-i)^l \bar V(l,n-l,m)
\end{equation}
which is the desired result. When dealing with the formalism of chiral
lagrangians it is more frequent to use amplitudes than $S$ matrix elements,
and so we will do from now on. It is straightforward then to rewrite our last
result as follows:

\begin{eqnarray}
\left( \prod_{i=1}^{n}\epsilon_{(L)\mu_i}\right)
T(\widetilde W^{\mu_1}_{a_1},..., \widetilde W^{\mu_n}_{a_n};A) = \hspace {7cm}
 \\ \nonumber
 = \sum_{l=0}^{n} (-i)^{n-l} \left( \prod_{i=1}^{l}v_{\mu_i} \right) \left(
\prod_{j=l+1}^{n}K^{a_j}_{\alpha_j} \right) \bar T(\widetilde W^{\mu_1}_{a_1}
...\widetilde W^{\mu_l}_{a_l},\omega_{\alpha_{l+1}} ...\omega_{\alpha_n};A)
\end{eqnarray}

  (notice that $l$ stands now for the number of gauge bosons, instead of the
number of GB).

	When one is dealing with amplitudes which satisfy the unitarity
bounds, one is sure that they will never grow with the energy and, due to the
fact that  $v_{\mu}\simeq O(M_{phys}/E)$, then it is
possible to neglect at high energies all terms in the RHS of eq.62 but that
with $l=0$, which is precisely the one where all the external $\widetilde W_L$
have been substituted by GB. Although these considerations are general, in
practice, the amplitudes are obtained through $\chi$PT as a truncated series in
powers of the energy. They satisfy unitarity just in the perturbative sense,
and
thus the same reasoning is not valid. We have to use power counting methods to
extract the leading contributions.

In our case, the power counting analysis goes as follows: We can, in
principle, make a formal Laurent expansion in $E/4\pi v$ of the amplitudes, but
 as they should satisfy the Low Energy Theorems (second reference in [3]) in
the
 $M^2 \ll E^2$ regime, we can always write the negative energy powers as
$(M/E)^{k}$. However, in the most interesting applications there will only be
even negative powers since they come from an even number of polarization
vectors
multiplied by the energy expansion of propagators. In practice, one typically
chooses the maximum positive power $N$ of the energy appearing in the
computations when fixing the maximum number of derivatives in the lagrangian
terms, so that we can write:
\begin{eqnarray}
 \bar T(\widetilde W^{\mu_1}_{a_1}
...\widetilde W^{\mu_l}_{a_l},\omega_{\alpha_{l+1}} ...\omega_{\alpha_n};A)
&\simeq&
\sum_{k=0}^{N} a_{l}^k \left(\frac{E}{4\pi v} \right) ^k + \sum_{k=1}^{\infty}
a_{l}^{-k} \left( \frac{M} {E}\right)^{k} \\ \nonumber
 ( \prod_{i=1}^{n}\epsilon_{(L)\mu_i} )
 T(\widetilde W^{\mu_1}_{a_1},..., \widetilde W^{\mu_n}_{a_n};A)
&\simeq& \sum_{k=0}^{N}
b^k  \left( \frac{E}{4\pi v} \right)^k + \sum_{k=1}^{\infty} b^{-k}
\left( \frac{M}{E} \right)^{k}
\end{eqnarray}

where we have set $g'=0$ momentarily. This series are formal and the
 coefficients $a^k_l,b^k$ (for the sake of brevity we have omitted their field
indices) could contain logarithms or further dependence in $g$. Therefore, in
addition to the derivative expansion we can now make perturbative calculations
on $g$ so that we can write $a_{l}^h=a_{lL}^h(1+ O(g/4\pi))$ where $a_{lL}^h$
is
the lowest order term in the expansion of $a^h_l$ in powers of $g$. Similarly,
and due to the fact that in most renormalization schemes we have $M \simeq
M_{phys} (1+O(g/4\pi) )$, we can expand $K$ too. That is:
$K^a_{\alpha}\simeq K^{a(0)}_{\alpha}+ K^{a(1)}_{\alpha} (g/4\pi)+....$.
 When these series are introduced in eq.62 we arrive to
 the following expression:
\begin{eqnarray}
 ( \prod_{i=1}^{n}\epsilon_{(L)\mu_i} )
T(\widetilde W^{\mu_1}_{a_1},..., \widetilde W^{\mu_n}_{a_n};A)  \simeq
\hspace{8cm} \\ \nonumber
\simeq \left(\prod_{j=1}^{n}K^{a_j (0)}_{\alpha_j} \right)
\sum_{k=0}^{N-n} (a_{0L}^{k}(1 + O(g/4\pi )))  \left( \frac{E}{4\pi v}
\right)^k
+O\left( \frac{M}{E} \right)+O\left( \frac{E}{4\pi v} \right)^{N-n+1}
\end{eqnarray}
 Where we have neglected terms of order $O(M/E)$ and $O(E/4\pi v)^{N-n+1}$, and
we have kept only the lowest order in the $g$ expansion of the energy
coefficients. This is the statement of the ET valid for the chiral lagrangian
description of the SM. Note that we are approximating the chiral expansion
coefficients of the amplitude with all the longitudinal gauge bosons, not by
the corresponding coefficients of the amplitude with all the $W_L$ substituted
by GB, but by their lowest order in the $g$ or $g'$ expansion. The validity of
the ET only to the lowest order in $g$ has been already suggested in the
literature (see the first cite in [14]), although in reference to the
complete amplitudes, not for each coefficient in the truncated chiral
expansions.

In order to check the last expression, we have explicitly
 computed the tree level amplitudes (so that all $Z$ factors are
 equal to one) up to four
derivatives obtained from the lagrangian of the second reference in [19] for
two processes, $Z_L^0 Z_L^0 \rightarrow Z_L^0 Z_L^0$ and $W_L^+W_L^-
\rightarrow Z_L^0Z_L^0$ using chiral  coordinates for the coset $S^3$. The
computation was done in two ways; first we have calculated the amplitude for
the
corresponding gauge bosons and we have proyected them into their
longitudinal components. Then those $S$ matrix were compared with the GB $S$
matrix elements to zeroth order in $g$ and $g'$ (see later the discussion
about $g'$ different from zero) and we found a perfect agreement at high
energies (above about $1 TeV$)
both analytical and numerically, for different
values of the chiral coupling constants (although of course, in the
large energy regime, the fourth derivative approximation is not expected
to be appropriate due to the bad energy behavior and the lack of
 unitarity of the amplitudes).

  Turning to the general case, and for practical purposes, if we want all the
approximations considered above to be reasonable, the allowed values of the
energy should fall into the following applicability window:
\begin{eqnarray}
 M \ll E \ll 4\pi v= 4 \pi M/g \\ \nonumber
 g/4\pi \ll (E/4\pi v)^{N-n+1}
\end{eqnarray}
The first two inequalities come from neglecting the $O(M/E)$ and $O(E/4\pi
v)^{N-n+1}$ terms respectively. The last constraint is needed to
reconcile both high and low energy requirements, since we are
 taking into account the $O(E/4\pi v)^{N-n}$ contribution while neglecting that
of
$O(M/E)$.

Now, the generalization to the case when $g' \neq 0$ is simple since $g' \ll g$
as
well as $ M_Z ^{phys} \simeq M_W ^{phys} \simeq M^{(a)}$ for any $a$ ( all the
different masses are of the same order when counting energy powers). Thus if we
keep the lowest order in the $g$ or $g'$ expansion of the $a$ coefficients the
approximation is still valid. Therefore, we can use eq.64 as the precise
statement of the ET for the $\chi$PT description of the SM.

	It is important to remark that this discussion in terms of energy expansions,
is due to the fact that at high energies the effective lagrangian does not
yield
a good unitary behavior for the truncated amplitudes (as it happens in
standard $\chi PT$), that could grow indefinitely [23]. That possibility does
not
allow us to consider only the $v_{\mu}=O(M/E)$ factors when extracting the
leading energy term in eq.62, since the amplitudes can contain positive powers
of $E$. However, if we could implement an unitary approximation, then the
unitarized amplitudes at high energies will never grow with a power
of $E$, and then we will be allowed to simply drop the terms with $v_{\mu}$
factors, thus obtaining:
\begin{equation}
 \left( \prod_{i=1}^{n}\epsilon_{(L)\mu_i}\right)
T(\widetilde W^{\mu_1}_{a_1},...,\widetilde W^{\mu_n}_{a_n};A)\simeq
\left(\prod_{j=1}^{n}K^{a_j}_{\alpha_j} \right)
T(\omega_{\alpha_1} ...\omega_{\alpha_n};A) + O(M/E)
\end{equation}
which is the usual formal statement of the ET. This unitarization procedure can
be achieved, for instance, by means of the Pad\'e approximants and dispersion
relations [23], large N-limit [24], etc..., and they could
enlarge considerably the ET applicability range.

Finally we would like to mention that another trivial example
 in which the version of the ET in eq.66 can be applied is in the MSM
considered
 as a particular case of the general models analysed in this work. This is so
since the effective action obtained once the Higgs field is
 formally integrated out, is a gauged NLSM of
the kind here considered, although all the couplings are now functions of the
Higgs mass (see for instance [25] for a computation of the coefficients
corresponding to the four derivatives terms). In addition, if one does not
integrate the Higgs field, the theory is renormalizable in the standard sense,
thus yielding a  good unitary and high energy behavior and therefore eq.66 can
be safely applied.

\section{Discussion}

In the previous section we finally arrived to the
version of the  ET valid for the gauged NLSM describing the GB and the gauge
boson interactions in the symmetry breaking sector
 of the SM. This theorem relates the $S$-matrix elements for the GB with those
of the longitudinal components of the GB $W_L$ at high energies. The
demonstration is based in the Ward-Slavnov-Taylor identities coming from the
BRS
symmetry of the quantized lagrangian and applies for any $R_{\xi}$-gauge and
for any choice of the coset coordinates or GB fields. The different
renormalization of the GB and the gauge fields appearing in the gauge fixing
condition produces some new factors that were absent in the original versions
of the ET.

	Still one
could ask about the real utility of the ET in the gauged NLSM or $\chi PT$
description of the symmetry breaking sector of the SM. In principle there is an
apparent contradiction in the terms
 of the title of this work, since $\chi PT$ provides a low-energy description
of
the GB dynamics as an expansion on the momenta over $4\pi v$ and the ET refers
to the large energy relation between the GB and the $W_L$'s $S$-matrix
elements.
Of course, the relevant point
 is whether there exists an applicability window for the theorem in the
intermediate energy region of eq.65 where  $\chi PT$ and the ET could be safely
applied simultaneously. For example, for the important case of the elastic
scattering of longitudinal gauge bosons the lowest bound obtained
from the second relation in eq.65 is equal to $1.7 TeV$ when the chiral
expansion is made considering terms with four derivatives. This suggests that
for the standard one-loop computation the ET applicability range would be
$1.7 TeV << E << 4\pi v \simeq 3.1 TeV$. This window is in principle narrow but
note that it is located in a energy region where one-loop $\chi PT$ is not a
very good approximation (for the case of the elastic pion scattering it
corresponds
roughly to $0.68 GeV << E << 1 GeV$).
In any case, only detailed computations of concrete
scattering processes can give us positive information about the existence,
the position and the
width of the window for the simultaneous application of standard $\chi PT$
and the ET.

 However, even when the
applicability range of, let say, one-loop  $\chi PT$
 and the ET were too  small to be useful, the description given here
 of the symmetry breaking sector of the SM  as a gauged NLSM and the
corresponding ET could
  be quite useful. This is because, apart from the perturbative one-loop (or
more) computations, there are non-perturbative procedures to extend the
$\chi PT$ to higher energies. These techniques include the use of the
dispersion
relations [23] and the large $N$ limit [24], $N$ being the number of GB. These
two methods work very well in the case of hadron physics (specially the first
one) and are also expected to do so in the case of the symmetry
breaking sector of the SM, where they could be combined safely with the ET.

Therefore we have in principle three different ways to apply the ET
depending on the kind of theory and the precise method used for making
computations: First we can consider the case of a renormalizable theory in
the standard sense with a finite number of parameters as, for example, the MSM.
In this case the good high energy behavior of the $S$ matrix elements is
granted and we arrive to the ET as stated in eq.66 without an upper energy
bound. For the practical purposes the only problem is the computation the the
GB $S$ matrix elements and the $K$ factors which depend on the
the gauge and the renormalization prescriptions. The
second scenario to be considered is when we use standard $\chi
PT$ to describe the SBS of the standard model and we truncate, as usual, the
chiral series  at some order in the number of derivatives. In
this case the precise statement of the theorem is that of eq.64 although only
in
the applicability window of eq.65. This version of the theorem is weaker, but
in
this case the computation of the $K$ factors is not so hard as in the previous
one  since we only need to know the lowest order in their $g$ and $g'$
perturbative expansion. Finally the SBS of the SM could also be described by
means of $\chi PT$ but making the computations in a non-standard way using some
kind of non-perturbative approximation like those discussed above so that we
have
a good high energy behavior of the $S$ matrix elements. In this case, the
version
of the ET is that of eq.66 without an upper energy bound or applicability
window  but including the $K$ factors.

\section{Conclusion}

 We will now briefly resume the main results of the present work:
 First, following the philosophy of $\chi PT$,
 we have built the gauged NLSM based on the
 coset $S^3=SU(2)_L \times SU(2)_R/SU(2)_{L+R}$ describing the GB and the gauge
boson dynamics of the symmetry breaking sector of the SM using all the
experimental information we have about this sector without further assumptions.
This description is done covariantly, i.e., it is independent of the
coordinates selected to parametrize the coset manifold and for the above
reasons it can be considered as the most general model independent description
of the SBS of the SM.

Then we have built the renormalized quantum lagrangian in the $R_{\xi}$-gauges
which are appropriate for doing perturbative computations. This
renormalized lagrangian is invariant under a set of renormalized (anti-) BRS
transformations and this fact can be used to find the corresponding
 Ward-Slavnov-Taylor identities for the renormalized Green functions, which
previously have been regularized dimensionally.

These Ward-Slavnov-Taylor identities were used to find a version of the ET
holding for the renormalized $S$-matrix elements. The final version of this
theorem has two important differences with the original one for the MSM: First,
it includes renormalization corrections which in general will depend on the
renormalization conditions and the gauge choice (the $K$ factors). Second, when
one works with the standard $\chi PT$ truncated energy expansions, it does not
 relate directly the
$S$-matrix elements with longitudinal gauge bosons to those with GB  but
 it relates the respective coefficients in both
expansions to the lowest order in $g$ or $g'$.

In principle, the energy window for the simultaneous application of
 $\chi PT$  at the one-loop level and the ET can be narrow but non-perturbative
methods could be used to extend it to higher energy regions thus providing a
model independent description of the symmetry breaking sector dynamics. We
sincerely believe that this kind of approach is the most sensible and
realistic, until no further information about
 this sector is available, and in particular, it can provide a very useful
phenomenological tool to  manage the precision test of the SM provided by LEP
and the future Large Hadron Collider (LHC) physics in an unified model
independent scheme.

\section{Acknowledgements}

This work has been supported in part by the
Ministerio de Educaci\'on y Ciencia (Spain)(CICYT AEN90-0034). We are very
grateful to M. Urdiales for helping us checking explicitly the ET, and
C.P.Martin
for helpful suggestions. A.D. also thanks the Gregorio del Amo Foundation
(Universidad Complutense)
for support and S.
Dimopoulos and  the Department of Physics
of the Stanford University for their kind hospitality.

\section{Note Added}

When this work was being completed we noticed the appearance of two
 related preprints on the subject. In [22] the authors arrive, using a
different  quantization procedure, to similar results to us for $g'=0$ and the
GB parametrization $U=exp(i\sigma^a \pi^a/v)$. They concentrate in the
renormalization factors which correct the ET version without the power counting
analysis. See also [26] for a discussion on this last issue.

\end{document}